# Nova explosions - The fascinating γ-ray emitting recurrent nova RS Ophiuchi

**Vincent Tatischeff[a], Margarita Hernanz[b,c,*]**

[a] *IJCLab (Université Paris-Saclay - CNRS/IN2P3), Bâtiment 104, 91405 Orsay Campus, France*
  *E-mail:* vincent.tatischeff@ijclab.in2p3.fr

[b] *Institute of Space Sciences (ICE-CSIC), Carrer de can Magrans, s/n, Campus UAB, 08193 Bellaterra (Barcelona), Spain*
  *E-mail:* hernanz@ice.csic.es

[c] *Institut d'Estudis Espacials de Catalunya (IEEC), Barcelona, Spain*

Classical and recurrent nova explosions occur on top of white dwarfs accreting H-rich matter from a companion main sequence or red giant star, in a close binary system. In the recent years, since the launch of the *Fermi* Gamma-Ray satellite by NASA in 2008, several novae have been detected by *Fermi*/LAT (LAT: Large Area Telescope) in High-Energy (HE) Gamma Rays, with energies larger than 100 MeV. This emission is known to be related to the acceleration of particles in the internal and/or external shocks occurring early after the thermonuclear nova explosion. However, Very-High-Energy (VHE) Gamma-Rays, with energies larger than 100 GeV, produced as a consequence of nova explosions have only been discovered very recently, in the recurrent nova RS Oph, that had an outburst in August 2021. These require the acceleration of protons, and not only of electrons; this was in fact predicted theoretically - based in observations at other wavelengths - in the previous eruption of RS Oph, in 2006, but has not been confirmed observationally until now. We review the origin of the different types of gamma-ray emission in novae and highlight the relevance of the recent VHE gamma-ray emission discoveries for the nova theory, mainly in the field of the mass ejection and the associated particle (electrons and protons) acceleration processes.

---

[*]Speaker

# 1. Introduction

White dwarfs (WDs) are the endpoints of stellar evolution of stars with masses smaller than about 10 $M_\odot$. WDs are degenerate stars made either of carbon and oxygen (CO WDs) or of oxygen and neon (ONe WDs), depending on the mass of their main sequence progenitor star. Their maximum mass is the Chandrasekhar mass (~1.4 $M_\odot$). Isolated WDs cool down to extremely low luminosities ($10^{-4}$–$10^{-5}$ $L_\odot$, for the older ones), whereas when they are in close binary systems and accrete matter from their companion star, they can host a thermonuclear explosion, either of hydrogen on their surface (nova explosion) or of carbon internally (Type Ia thermonuclear supernova explosion).

WDs can accrete matter and subsequently explode in mainly two types of binary systems. The first is a cataclysmic variable, where the companion of the WD is a main sequence star, similar to the Sun. There, mass transfer occurs via Roche lobe overflow. The typical orbital periods are hours-days. Hydrogen burning in degenerate conditions on top of the WD leads to a thermonuclear runaway and a (classical) nova explosion. The amount of mass ejected is uncertain, but typically it is about $10^{-4}$–$10^{-5}$ $M_\odot$, with velocities of 100's or 1000's of km/s. A nova explosion does not disrupt the WD, contrary to what occurs in thermonuclear - Type Ia - supernova explosions. Therefore, after enough mass is accreted again from the companion star, a new explosion will occur, with typical recurrence time of $10^{(4-5)}$ years. Another scenario is the so-called symbiotic binary, where the WD accretes matter from the stellar wind of a red giant companion. Typical orbital periods of these systems are a few 100 days, larger than in cataclysmic variables. This scenario leads to more frequent nova explosions than in cataclysmic variables, being the recurrence periods shorter than 100 years, and therefore more than one outburst can have been recorded. Recurrent novae can in fact also happen in binary systems without a red giant companion. There are about 10 known recurrent novae in the Galaxy [1] and a few others in nearby galaxies, like M31 (Andromeda galaxy) or in the Magellanic Clouds. The best estimate of the Galactic nova rate is $50^{+31}_{-23}$ yr$^{-1}$, for novae of any type [2].

Recurrent novae are possible scenarios of type Ia supernovae, because they are expected to occur in massive WDs, close to the Chandrasekhar mass limit. However, the absence of H in the spectra of SNIa is difficult to explain if the exploding WD is accreting matter in a cataclysmic variable or in a symbiotic binary system.

The recurrent nova RS Ophiuchi (RS Oph) is a very well observed and studied recurrent nova, within a symbiotic binary system, with an accreting WD that has an estimated mass between 1.2 and 1.4 $M_\odot$, a red giant donor with mass in the range 0.68−0.80 $M_\odot$ and orbital period 453.6 ± 0.4 days [3]. There have been several recorded outbursts of RS Oph, being the latest ones in 1985, 2006 and 2021; the approximate recurrence time is 15 years, ranging between 9 and 27 years [1]. See [4] for a complete description of novae and [5] for a dedicated study of the RS Oph recurrent nova during its 2006 eruption.

RS Oph has been detected at almost all energies, and excellent data has been obtained, especially in its two latest eruptions in 2006 and 2021. In particular - as in fact we predicted in 2007 before *Fermi* launch (see [6], and also [7,8]) - RS Oph was detected with *Fermi*/LAT in HE γ-rays, with $E$ ~ (0.1-10) GeV, during its 2021 eruption [9].

Other novae have also been detected with *Fermi* LAT since its launch in 2008: see review about classical novae [10], which puts special emphasis on the role played by shocks on particle acceleration in novae, leading to HE γ-ray emission; a summary of the *Fermi*/LAT detections of novae up to 2020 is presented in [10]. However, no nova had been detected in VHE γ-rays before the RS Oph eruption in 2021, when both MAGIC [11] and H.E.S.S. [12] detected it. In this paper we explain our model of cosmic-ray acceleration in RS Oph and its relationship with the recent HE and VHE γ-ray detections of this recurrent nova.

## 2. Model of cosmic-ray acceleration in RS Oph

Cosmic rays (CRs) accelerated in supernova shocks can have observable impacts on the remnant morphology [13], the post-shock temperature [14] and the amplified magnetic field in the shock region [15]. These effects are well explained by the theory of nonlinear diffusive shock acceleration (NLDSA), which predicts that the backpressure exerted by accelerated energetic ions can strongly modify the shock structure [16, 17]. Evidence for NLDSA in the 2006 outburst of RS Oph came from two observations [6]. First the post-shock temperature measured in X-rays was lower by a factor of ~3 than that given by the usual relation for a test-particle strong shock: $kT_S = (3/16)\,\mu m_H\, V_S^2$, where $k$ is the Boltzmann constant, $\mu m_H$ the mean particle mass and $V_S$ the forward shock velocity, which was estimated from IR spectroscopic observations. In addition, both IR and X-ray observations showed that the shock has decelerated at a higher rate than predicted by the test-particle adiabatic shock wave model. Noteworthy, the characteristic timescale of shock evolution is ~$10^5$ shorter for RS Oph outbursts than for supernovae, for which it is not possible to study the time dependence of the shock structure.

In [6], we have shown that these two observations could be explained by NLDSA of protons with a moderate acceleration efficiency characterized by a proton injection rate into the acceleration process of $\eta_{inj} \sim 10^{-4}$. We predicted that accelerated protons achieved a maximum energy of the order of a few TeV in less than a week after outburst. In [7] (see also [8]), using X-ray photoelectric absorption measurements to estimate the density of the red giant wind as a function of radius, we then calculated the γ-ray emission from neutral pion ($\pi^0$) production and found it to be above the detection sensitivity of *Fermi* LAT (*Fermi* was not yet launched in 2006). In Figure 1, we compare our predicted γ-ray light curve to the one observed by the LAT during the 2021 outburst of RS Oph [9]. We see that beyond day 1 – the assumed onset of particle acceleration in the [6] model – the agreement of the prediction with the data is remarkable.

### 2.1 Insights from the γ-ray observations of RS Oph 2021

We now discuss the [6] model in light of the recent observations of RS Oph 2021.

#### 2.1.1 Source distance

Parallax measurements with *Gaia* give a source distance of $2.4^{+0.3}_{-0.2}$ kpc [18], but the reliability of the *Gaia* measurements has been questioned due to the orbital motion in RS Oph [19]. However, the *Gaia* result agrees well with the estimate of [20], $D = 2.45 \pm 0.4$ kpc, which is based on high-resolution VLBA radio imaging of RS Oph 2006. It was obtained by comparing the angular size of the observed expanding shell of synchrotron emission with the shock radius $R_S$ inferred from optical and IR spectroscopic observations. We used the distance from [20] in our γ-ray emission model of RS Oph 2006 [7], as did the MAGIC team in their analysis of RS Oph 2021 [11]. But the H.E.S.S. and *Fermi* LAT papers assumed a shorter distance, $D = 1.4$ kpc [12] or 1.6 kpc [9], based on previous estimates (e.g., [21]). This explains part of the difference in the derived CR acceleration efficiency (Sect. 2.1.5).

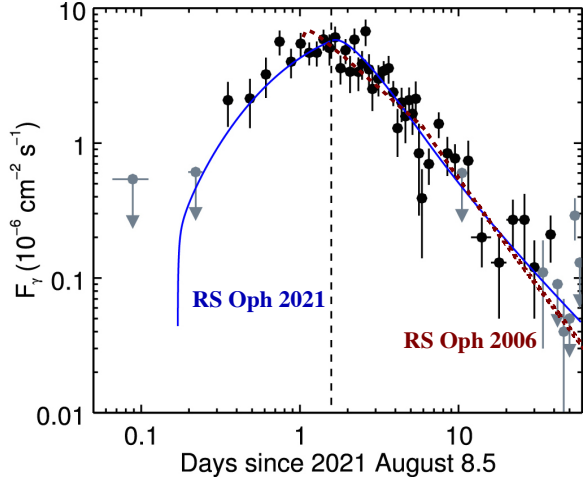

Figure 1: Predicted γ-ray light curve of RS Oph 2006 at energies > 100 MeV (red dashed line, from [7]; see also [8]) compared to the γ-ray light curve of RS Oph 2021 observed by *Fermi* LAT [9]. The blue solid line shows the hadronic model from the LAT paper. Figure adapted from [9].

### 2.1.2 Maximum energy of the accelerated protons

The maximum proton energy predicted by [6] is shown in Figure 2. It depends on three poorly known parameters: the strength of the amplified magnetic field in the shock upstream ($\alpha_B$), the value of the particle diffusion coefficient relative to the Böhm limit ($\eta_{\rm mfp}$), and the position of the particle free escape boundary ahead of the shock in units of the shock radius ($f_{\rm esc}$). H.E.S.S. data for RS Oph 2021 suggest that protons reached a maximum energy of the order of 10 TeV ~ 4–5 days after outburst (measured maximum γ-ray energy of ~1 TeV). Such a high maximum energy suggests that plasma turbulence at the shock was strongly amplified by CR-driven instabilities up to the saturation level, leading to a high turbulent magnetic field ($\alpha_B$) and a low diffusion coefficient ($\eta_{\rm mfp}$); see also [12] for a discussion on $E_{\rm max}$. However, the maximum proton energies (in the sense of an exponential cut-off) estimated from the MAGIC data [11] suggest a weaker B-field amplification (see Figure 2).

### 2.1.3 Hadronic origin of the γ-ray emission

In [7] (see also [8]), we compared the γ-ray luminosity expected from $\pi^0$ production to that from inverse Compton (IC) scattering of shock-accelerated electrons. We estimated the latter from the luminosity of the nova ejecta $L_{\rm ej}$ and the non-thermal synchrotron luminosity $L_{\rm syn}$ observed by [22] at radio frequencies lower than 1.4 GHz: $L_{\rm IC} = L_{\rm syn} \times U_{\rm rad}/(B_{\rm ps}^2/8\pi)$, where $U_{\rm rad} \sim L_{\rm ej}/(4\pi c R_S^2)$ and $B_{\rm ps}$ is the post-shock magnetic field ($c$ is the speed of light). The ejecta luminosity was taken to be the Eddington luminosity, $L_{\rm ej} \sim 10^{38}$ erg s$^{-1}$. We then found the γ-ray luminosity from the leptonic contribution to be about two orders of magnitude lower than that from $\pi^0$ production.

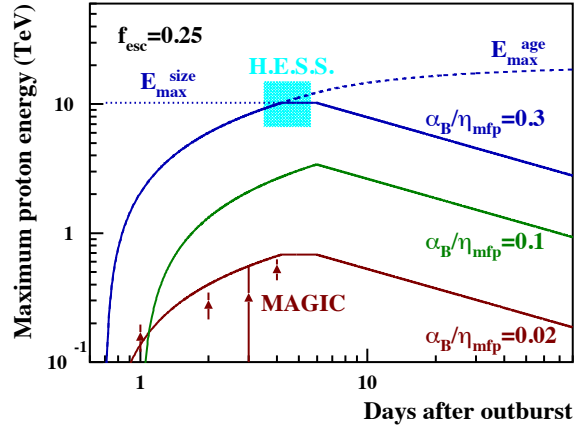

Figure 2: Maximum proton energy as a function of time after outburst. *Green line*: from [6] for RS Oph 2006; *blue lines*: calculated with $\alpha_B/\eta_{mfp} = 0.3$ so that it reaches ~10 TeV at days ~ 4 – 5 as suggested by the H.E.S.S. observations of RS Oph 2021 [12]; *red line*: with $\alpha_B/\eta_{mfp} = 0.02$, which provides a better agreement to the MAGIC data [11]. $E_{max}^{age}$ and $E_{max}^{size}$ are the maximum energies caused by the finite shock age and size, respectively. The parameters $\alpha_B$, $\eta_{mfp}$ and $f_{esc}$ are discussed in the text.

The hadronic origin of the γ-ray emission detected during the 2021 outburst is favoured for three main reasons. First, the measured curvature of the γ-ray spectrum between 50 MeV and 250 GeV is better described by the hadronic model than by the leptonic one [11]. Second, the inferred increase of the particle maximum energy during the first four days after outburst is difficult to explain in the leptonic model, which predicts a fast IC cooling of high-energy electrons. Third, the leptonic model requires that a significant fraction of the shocked gas internal energy is transferred to non-thermal electrons, $f_e > 1\%$, which is not expected in the DSA theory (see [12]).

### 2.1.4 Slope of the accelerated proton spectrum

The γ-ray spectra of RS Oph 2021 were modelled with a hadronic scenario in both the H.E.S.S., MAGIC and *Fermi*-LAT papers, assuming for the proton injection spectrum either a power-law in energy or in momentum. The fitted slope of the high-energy proton spectrum was found to be $s_p \sim -2.2$, which is steeper than the test-particle prediction of DSA: $s_p = -2$. However, recent developments of the NLDSA theory predict that the particle spectrum becomes steeper than the test-particle prediction when sufficient particles are injected into the DSA process, due to the drift of CR-generated magnetic perturbations [17, 23, 24]. Efficient particle acceleration leads to strong magnetic field amplification due to CR-streaming instabilities and the velocity of the generated scattering centres relative to the background plasma makes the fast particles feel an effective compression factor lower than 4, thus the steeper particle spectrum.

Figure 3 shows calculated phase-space distributions of shocked protons at day 4 after outburst (the phase-space distribution function $f(p)$ is related to the energy spectrum $N(E)$ by $N(E) = 4\pi p^2 f(p) dp/dE$), using the NLDSA model originally developed by [6], but since extended to take into account magnetic field amplification from the resonant streaming instability, the back-reaction of the magnetic pressure on the shock structure [25], as well as the Alfvénic drift effect leading to steeper particle spectra [17]. With the proton injection rate deduced by [6] from the measured post-shock temperature, $\eta_{\rm inj} \sim 1.4 \times 10^{-4}$, the non-thermal proton distribution is found to be flatter than the one deduced from the γ-ray observations of RS Oph 2021, which suggests that B-field amplification by the resonant streaming instability is not enough to account for the observations. To illustrate that a stronger magnetic field amplification could explain the inferred slope of the proton spectrum, we show in Figure 3 a spectrum obtained by arbitrarily multiplying by a factor of 10 the upstream magnetic pressure expected from the resonant streaming instability. In fact, [23] and [24] recently found that the steepening of accelerated particle spectra at strong shock could mainly be due to the advection downstream of strong magnetic perturbations excited by the non-resonant Bell instability. This model could well explain a slope of $f(p)$ of -4.2 (= $s_p$-2) for a shock speed of 4000–5000 km/s (see Fig. 1 of [24]).

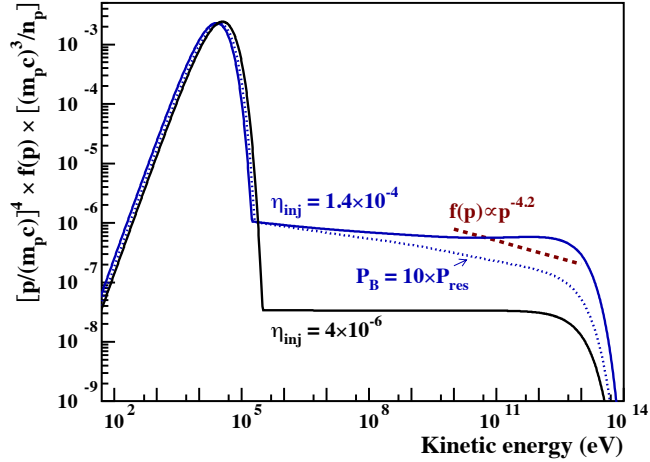

Figure 3: Shocked proton phase-space distributions vs. kinetic energy at day 4 after outburst of RS Oph 2006 for two values of the proton injection rate $\eta_{\rm inj}$. The distribution functions have been multiplied by $[p/(m_{\rm p}c)]^4$ to flatten the spectra and by $[(m_{\rm p}c)^3/n_0]$ to make them dimensionless (here $p$, $m_{\rm p}$ and $n_{\rm p}$ are the proton momentum, mass and number density ahead of the shock). The blue dotted line shows the distribution obtained by multiplying the upstream magnetic pressure expected from the resonant streaming instability by an order of magnitude. The red dashed line illustrates a slope of -4.2 as deduced from *Fermi* LAT, MAGIC and H.E.S.S. observations (see text).

We also show in Figure 3 a phase-space distribution function calculated for $\eta_{\rm inj} = 4 \times 10^{-6}$, which is the proton injection rate obtained by [9] in their modelling of the *Fermi* LAT data. No significant amplification of the magnetic field and steepening of the proton spectrum can be expected for such a low injection rate, for which the test particle approximation applies.

**2.1.5 Cosmic-ray acceleration efficiency**

The low value of $\eta_{\rm inj}$ found by [9] would imply that only ~1% of the shock kinetic energy was transferred to accelerated protons in the 2021 outburst of RS Oph, which is significantly lower than the acceleration efficiency of ~20% found in ref. [11], the latter result being in good agreement with the predictions of [6]. A value of $\eta_{\rm inj} = 4 \times 10^{-6}$ does not lead to observable non-linear effects of particle acceleration, such as those we evidenced in our 2007 paper. The difference between [9] and [7] in the acceleration efficiency required to explain the γ-ray light curve (see Figure 1) can partly be explained by the lower shock velocity considered in the *Fermi*-LAT paper. The authors of this paper used Swift-XRT temperature measurements to estimate the shock speed, and thus found values of $V_S$ lower than that used by [7] by a factor of ~1.7 (see Sect. 2). Another difference between the models of [9] and [7] comes from the source distance, taken to be $D$ = 1.6 kpc in [9] instead of 2.45 kpc in [7]. As $\eta_{inj}$ varies as $V_S^3 \times D^2$, these two effects can explain an order of magnitude difference the derived injection rates.

## 3. Summary and conclusions

Symbiotic recurrent novae behave like "miniature" supernova remnants, evolving about $10^5$ times faster and being much dimmer. Nearly forty eruptions of the ten known galactic recurrent novae have been recorded since the beginning of the 20[th] century [1]. The next eruption could be that of the nova T CrB, which is located at only 0.8 kpc from Earth and is recently showing an increasing activity [26].

RS Oph 2021 is the first nova outburst to be detected at VHE γ-rays. The predictions we made after the 2006 outburst of a proton acceleration beyond TeV energies and a detectable γ-ray emission from hadronic interactions have been remarkably confirmed by the ground-based and space-based γ-ray observations of the 2021 outburst (see Figure 1). The very similarity of the shock evolution in the two most recent outbursts of RS Oph is evident from the data of the *Neil Gehrels Swift Observatory*, which monitored in detail the two eruptions [27]. Multi-wavelength observations of both outbursts revealed several effects related to an efficient particle acceleration at the blast wave: shock deceleration at a higher rate than that expected from the radiative losses, reduced post-shock temperature compared to the test-particle prediction, strong magnetic field amplification, significant steepening of the accelerated proton spectrum. No other astrophysical object allows us to observe all these non-linear effects of particle acceleration at the same time. RS Oph is certainly a key laboratory to study the microphysics of particle acceleration in astrophysical shock waves.

**Acknowledgements**: MH acknowledges funding from MICIN/AEI grant PID2019-108709GB-I00 and program Unidad de Excelencia Maria de Maetzu CEX2020-001058-M.